\documentclass[aps,reprint,nofootinbib,twocolumn,superscriptaddress,longbibliography]{revtex4-2}
\usepackage{bbold}
\usepackage{physics}
\usepackage{comment}
\usepackage{natbib}
\usepackage{slashed}

\usepackage[english]{babel}
\usepackage{graphicx,epstopdf}
\usepackage{amssymb,amsmath}
\usepackage{dcolumn}
\usepackage{bm}
\usepackage{color}
\usepackage{enumerate}
\usepackage[colorlinks=true,citecolor=blue]{hyperref}
\hypersetup{colorlinks=true,citecolor=blue,linkcolor=blue,urlcolor=blue}
\usepackage{soul}
\usepackage{tabularx}
\usepackage[dvipsnames]{xcolor}

\begin{document}

\preprint{APS/123-QED}

\title{Disorder-Induced Topological Transitions in a Multilayer Topological Insulator}

\author{Z.Z. Alisultanov}
\email{zaur0102@gmail.com}
\affiliation{Abrikosov Center for Theoretical Physics, MIPT, Dolgoprudnyi, Moscow Region 141701, Russia}
\affiliation{Institute of Physics of DFRS, Russian Academy of Sciences, Makhachkala, 367015, Russia}
\author{A. Kudlis}
\email{andrewkudlis@gmail.com}
\affiliation{Abrikosov Center for Theoretical Physics, MIPT, Dolgoprudnyi, Moscow Region 141701, Russia}
\affiliation{Russian Quantum Center, Skolkovo, Moscow, 121205, Russia}


\begin{abstract}We examine the impact of non-magnetic disorder on the electronic states of a multilayer structure comprising layers of both topological and conventional band insulators. Employing the Burkov-Balents model with renormalized tunneling parameters, we generate phase diagrams correlating with disorder, demonstrating that non-magnetic disorder can induce transitions between distinct topological phases. The subsequent section of our investigation focuses on the scenario where disorder is unevenly distributed across layers, resulting in fluctuations of the interlayer tunneling parameter - termed off-diagonal disorder. Furthermore, we determine the density of states employing the self-consistent single-site diagram technique, expanding the Green function in relation to the interlayer tunneling parameter (locator method). Our findings reveal that off-diagonal disorder engenders delocalized bulk states within the band gap. The emergence of these states may lead to the breakdown of the anomalous quantum Hall effect (AQHE) phase, a phenomenon that has garnered significant attention from researchers in the realm of topological heterostructures. Nonetheless, our results affirm the stability of the Weyl semimetal phase even under substantial off-diagonal disorder.

\end{abstract}

\maketitle


\section{Introduction}
\allowdisplaybreaks
The study of topological materials continues to be a “hot” direction in the modern theory of condensed matter~\cite{Hasan,Qi,Armitage,Anirban,Tokura}. One of the most pressing questions is the influence of various types of disorder on topological phases~\cite{Liu2009,Groth,Guo,Meier}. Disorder plays a key role in establishing certain properties of electronic systems. For example, disorder is the cause of quantum localization (Anderson localization), as well as Hall plateaus in the integer quantum Hall effect. Another interesting effect is disorder-induced topological phases in two- and three-dimensional systems, which have been called two- and three-dimensional Anderson topological insulators. Disorder in these systems, on the one hand, induces Anderson localization, and on the other hand, the topological mass is renormalized due to disorder until the sign changes (band inversion) and the system undergoes transition to a topologically nontrivial state~\cite{Groth}. Thus, disorder induces chiral edge states in the system, which in the pure state is topologically trivial. The disorder-induced metal-insulator transition has also been studied in Weyl semimetals (see, for example, ~\cite{Park,Zhang2022}). In recent work, these effects were studied in a two-layer structure~\cite{Krishtopenko}. It is worth noting that the topological Anderson phase was predicted not only for crystalline systems, but also for amorphous ones (see, for example, ~\cite{XiaoyuCheng}).

In this work, we study the effects of nonmagnetic disorder in a multilayer system of topological (TI) films and conventional (normal) insulators within the Burkov-Balents analytical model~\cite{Burkov}. The study of various TI heterostructures is of particular interest due to the emergence of unique effects in such systems~\cite{Fan,Kandala,Hesjedal,Chong,Eremeev,Liu,Tokura19,Bernevig,Jiang}. One such effect is the anomalous quantum Hall effect (AQHE), and many ongoing works are aimed at searching for signs of the topological magnetoelectric effect and axion insulator phases. In a multilayer structure, the AQHE mode can be achieved due to the proximity effect without introducing magnetic impurities directly into the TI layers, it means that magnetic impurities are introduced into the layers of a conventional insulator. Thus, the affect only the spin degeneracy of edge modes (which is necessary to obtain the AQHE and Weyl phase modes), but are not the centers of scattering of these modes. Another fundamental motivation for studying multilayer topological systems is related to the fact that such structures are convenient for observing phenomena associated with the quantum geometry (see, for example, ~\cite{Törmä,Gao}). This concept is the most general and fundamental approach to topological condensed matter physics. In PT-symmetric heterostructures (this is achieved by appropriate selection of layers), in which the Chern number is identically equal to zero due to symmetry, the so-called nonlinear quantum Hall effect has been measured~\cite{Gao}. The origin of this effect can be explained by the nontrivial quantum metrics of Bloch bands. On the other hand, recent work has investigated multilayer magnetic topological insulators with an asymmetric layered structure, which makes it possible to achieve a magnetic field-controlled phase transition to the AQHE regime.

The paper is organized as follows. In Sec.~\ref{sec:II} we recall the main aspects of the Burkov-Balents model for multilayer TI. In Sec.~\ref{sec:III} we examine the disorder-induced renormalization of the edge modes and tunnelling parameters of the model used and plot phase diagrams for various cases as a function of disorder. In Sec.~\ref{sec:IV}, we investigate the density of states for the situation where disorder is unevenly distributed across layers. This leads to the fact that the interlayer tunnelling parameter randomly changes (fluctuates) from layer to layer -- off-diagonal disorder. At the we draw a conclusion, which briefly summarizes the main results.

\section{Burkov-Balents model}
\label{sec:II}

\begin{figure}[t]
\includegraphics[width = 1.0\linewidth]{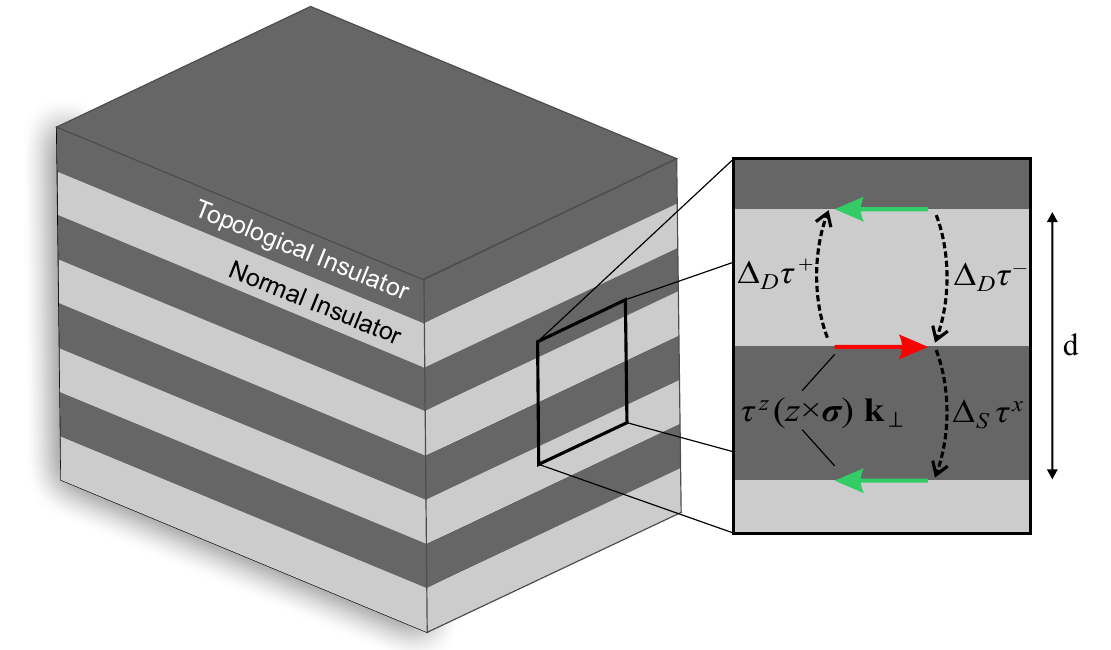}
\caption{Schematic representation of the system under study. Horizontal arrows on the right panel represent chiral edge modes in the TI films. Different colors of arrows (green and red ones) correspond to different chiralities. A unit cell of height $d$ contains one TI layer and one normal insulator layer.}
\label{fig:mulyilayer model}
\end{figure}

The multilayer system of conventional insulators and TIs we are studying is shown in Fig.~\ref{fig:mulyilayer model}. To study such a system we use the well-known Burkov-Balents model, which is quite effective for studying topological properties. The Hamiltonian of the system within this model can be written in the following form~\cite{Burkov}
\begin{multline}
\mathcal{H}=\sum_{\mathbf{k}_{\perp},i,j}c_{\mathbf{k}_{\perp}i}^{\dagger}\left[\left(\tau^{z}\upsilon_{F}\left(\hat{\bm{z}}\times\bm{\sigma}\right)\cdot\mathbf{k}_{\perp}+\Delta_{S}\tau^{x}\right)\delta_{i,j}\right. \\
\left.+\frac{1}{2}\Delta_{D}\tau^{+}\delta_{j,i+1}+\frac{1}{2}\Delta_{D}\tau^{-}\delta_{j,i-1}\right]c_{\mathbf{k}_{\perp}j},
\end{multline}
where $\upsilon_{F}\left(\hat{\bm{z}}\times\bm{\sigma}\right)\cdot\mathbf{k}_{\perp}\tau^{z} $
is the Hamiltonian of the edge modes of the TI film, in which there are two chiralities are given by the Pauli matrix $\tau^{z}$, $\mathbf{k}_{\perp}=\left(k_{x},k_{y}\right)$. The Pauli matrices $\bm{\tau}$ define the degrees of freedom associated with the top and bottom edges of the TI layer, and the Pauli matrices $\bm{\sigma}$ describe the spin degrees of freedom. The indices $i,j$ number the TI layers. The term with $\Delta_{S}$ matches tunneling between edge modes inside the TI film (i.e., hopping between the upper and lower surfaces of the same layer), and the terms with $\Delta_{D}$ correspond to tunneling between edge modes neighboring films (see Fig.~\ref{fig:mulyilayer model}). In the momentum representation, this Hamiltonian has the following form~\cite{Zaur}

\begin{equation}
\mathcal{H}_{\mathbf{k}}=\left(\begin{array}{cc}
\upsilon_{F}\left(\hat{\bm{z}}\times\bm{\sigma}\right)\cdot\mathbf{k}_{\perp} & \Delta_{S}+\Delta_{D}e^{ik_{z}d}\\
\Delta_{S}+\Delta_{D}e^{-ik_{z}d} & -\upsilon_{F}\left(\hat{\bm{z}}\times\bm{\sigma}\right)\cdot\mathbf{k}_{\perp}
\end{array}\right),
\end{equation}
with the spectrum
\begin{equation}
\varepsilon_{k}=\pm\sqrt{\upsilon_{F}^{2}\left(k_{x}^{2}+k_{y}^{2}\right)+\Delta^{2}\left(k_{z}\right)},
\end{equation}
where the function $\Delta$ is  defined as $\Delta^{2}\left(k_{z}\right)=\Delta_{S}^{2}+\Delta_{D}^{2}+2\Delta_{S}\Delta_{D}\cos k_{z}d$.

Further, without loss of generality, we assume that $\Delta_{S},\Delta_{D}>0$. Let us note two features of this Hamiltonian that are important in the rest of the paper. First, for $\Delta_{S}=\Delta_{D}$ the system is gapless and contains the Dirac point $\left(0,0,\pi/d\right)$. Secondly, this Dirac point in fact can be considered  as a critical point of a  phase transition between topological ($\Delta_{S}<\Delta_{D}$) and ordinary ($\Delta_{S}>\Delta_{D}$) insulators~\cite{Murakami}. This transition is described by the topological invariant $\nu$, which is determined from the expression $(-1)^{\nu}=\text{sgn}(\Delta_{S}-\Delta_{D})$. For other details of the model, the reader can refer to the work~\cite{Burkov}.

\section{Disorder-induced renormalization of edge modes and phase transitions in multilayer TI}
\label{sec:III}

Let us now perform the generalization of the Burkov-Balents model to the case of the presence of non-magnetic disorder inside the TI layers. First of all, we briefly discuss the effect of such disorder on the electronic states of an isolated TI film. The main idea of our work is based on the fact that the disorder inside the TI film rearranges its edge modes, tunneling between which ultimately determines the electronic structure of the multilayer structure. In this section, we consider in detail the renormalization of TI edge states and the consequences of such renormalization in the form of topological phase transitions.

Following the works~\cite{Qi,Hosur,Guo}, we use for TI with disorder the following 4-band Hamiltonian for a cubic lattice (for simplicity, the lattice constant is set equal to unity)
\begin{equation}\label{eq:disordered_Hamiltonian}
H_{TI}\!=\!\sum_{\mathbf{k}}\!\varPsi_{\mathbf{k}}^{\dagger}\!\left[\sum_{\mu=0}^{3}\!\!d_{\mu}\!\left(\mathbf{k}\right)\!\Gamma_{\mu}\!+\!d_{4}\!\left(\mathbf{k}\right)\!\mathbb{I}\!\right]\!\varPsi_{\mathbf{k}}\!+\!\sum_{j}\!U_{j}\varPsi_{j}^{\dagger}\varPsi_{j},
\end{equation}
where $\varPsi_{j}$ is a 4-component state vector at the $j^{\textup{th}}$ site of the crystal lattice, $d_{0}\left(\mathbf{k}\right)=\chi-2t\sum_{i }\cos k_{i}$, $d_{i}\left(\mathbf{k}\right)=-2\lambda\sin k_{i}$, $d_{4}\left(\mathbf{k }\right)=2\gamma\sum_{i}\left(1-\cos k_{i}\right)$, $\Gamma_{\mu}$ are Dirac matrices, $U_{j}$ is a random potential at the $j^{\textup{th}}$ lattice site caused by disorder. We use the simplest Anderson model, within which the energy values at the sites are distributed uniformly with a density of $1/U_{0}$ in the interval $\left[-U_{0}/2,U_{0}/2\right]$.

For $U_{0}=0$ we obtain the Hamiltonian of the pure system $H_{TI}^{0}=\sum_{\mathbf{k}}\varPsi_{\mathbf{k}}^{\dagger}H_{\mathbf{k}}^{0}\varPsi_{\mathbf{k}}$. This Hamiltonian contains a term $m=\chi-6t$, called the topological mass. It can be shown that for $m>0$ the system is an ordinary band insulator with a gap value equal to $m$. If $m<0$, then the spectrum becomes inverted and the system, in addition to gapped bulk states, also contains gapless edge states that have topological protection. This phase is called a TI. Note that for $U_{0}=0$ the Hamiltonian is invariant under time reversal, since $\mathcal{T}\Gamma_{0}\mathcal{T}^{-1}=\Gamma_{0}$, $\mathcal{T}\Gamma_{i}\mathcal{T}^{-1} =-\Gamma_{i}$, $d_{i}\left(-\mathbf{k}\right)=-d_{i}\left(\mathbf{k}\right)$, where $\mathcal{ T}$ is the time inversion operator. This invariance is the basis for topological protection of chiral edge states.

\begin{figure}[t]
\includegraphics[width = 0.6\linewidth]{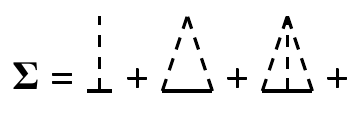}
\caption{Diagram series for the self-energy part (\ref{eq:self-energy}) in the single-impurity
cross technique in the T-scattering matrix approximation.}
\label{fig:diagram1}
\end{figure}

At $U_{0}\neq0$ the spectrum of the system is renormalized. Within the single-impurity diagram technique~\cite{abrikos1,abrikos2,edward} the self-energy part can be depicted as in Fig.~\ref{fig:diagram1}. The use of such a diagrammatic series means going beyond the Born approximation. As we see below, this eliminates some singularities from the calculations and gives more accurate results. The Born approximation can be applied if the condition is met $k_{F}^{3}\int u\left(\bm{r}\right)d^{3}\mathbf {r}\ll\epsilon_{F}$ (see, for example, Ref.~\cite{abrikos3}). If we assume that the impurity potential is equal to $u_{0}$ in the region $\Delta^{3}\mathbf{r}$ and is close to zero in the rest of the space, then this criterion can be rewritten as $\Delta^{3 }\mathbf{r}\ll\epsilon_{F}/u_{0}k_{F}^{3}\simeq 1/\left(u_{0}\left(2m\right)^{3/2}\epsilon_{F}^{1/2}\right)$. Thus, the criterion is satisfied for a short-range potential and small values of the Fermi energy and topological mass. This is important because the transition from a topologically trivial phase with $m>0$ to a topologically nontrivial phase with $m<0$ is carried out through the value $m=0$. It should also be noted that we do not take into account diagrams with intersections from several impurities. The smallness of such diagrams is estimated as $\hbar/\epsilon_{F}\tau\gg1$~\cite{abrikos3}. This condition can be interpreted as follows: the shorter the relaxation time $\tau$, the larger the diffusion volume covered by the particle and the lower the probability of the particle returning to the starting point. We assume that the both mentioned criteria (the applicability of the Born approximation and neglecting intersected diagrams) are satisfied in our system. Thus, the diagrammatic series in Fig.~\ref{fig:diagram1} can be summed in the short-range impurity center approximation, when the Fourier transform of the impurity potential can be considered as  a  constant. In this case
\begin{align}\label{eq:self-energy}
\hat{\Sigma}\left(\epsilon_{F}\right)=U\left(1-U\sum_{BZ}\hat{G}\left(\epsilon_{F},\mathbf{k}\right)\right)^{-1},
\end{align}
where $\hat{G}_{\mathbf{k}}\left(\epsilon_{F}\right)=\left(\epsilon_{F}+i\delta-H_{\mathbf{k}}^{0}-\Sigma\right)^{-1}$. Averaging over impurities gives
\begin{equation}
\hat{\Sigma}=-\frac{\left(\hat{F}^{-1}\right)^{2}}{U_{0}}\ln\frac{1-\frac{U_{0}}{2}\hat{F}}{1+\frac{U_{0}}{2}\hat{F}}-\hat{F}^{-1},
\label{eq:self-energy averaging}
\end{equation}
where we  introduce the following notation: $\hat{F}=\sum_{BZ}\hat{G}_{\mathbf{k}}$ and $\hat{F}^{-1}\hat{F}=\mathbb{I}$. The matrix structure of the Hamiltonian leads to a similar structure of the self-energy part. Therefore, the latter can be expanded in the basis of Dirac matrices $\Sigma=\sum_{\mu}\Gamma_{\mu}\Sigma_{\mu}+\mathbb{I}\Sigma_{4}$, where $\Sigma_{\ mu}=1/4\text{tr}\left(\Gamma_{\mu}\Sigma\right)$, $\Sigma_{4}=1/4\text{tr}\Sigma$. We consider the case of non-magnetic impurities. This means that the total Hamiltonian~\eqref{eq:disordered_Hamiltonian} has to be also $T$-invariant. Thus, $\mathcal{T}\Sigma\mathcal{T}^{-1}=\Sigma$ or $\mathcal{T}\Sigma_{i}\mathcal{T}^{-1}=-\Sigma_{i}$. However, the quantities $\Sigma_{\mu}$ are functions of energy only and do not depend on momentum. Therefore $\mathcal{T}\left(d_{i}\left(\mathbf{k}\right)+\Sigma_{i}\right)\mathcal{T}^{-1}=-d_{i} \left(\mathbf{k}\right)+\Sigma_{i}$  should be true. If we take into account that $\mathcal{T}\Gamma_{i}\mathcal{T}^{-1}=-\Gamma_{i}$, then we finally obtain $\Sigma_{i}\equiv0$. Thus, non-magnetic impurities lead to renormalization of the topological mass and Fermi energy due to the quantities $\Sigma_{0}$ and $\Sigma_{4}$, respectively: $m\rightarrow\overline{m}=m+\text{Re}\left[\Sigma_{0}\right]$, $\epsilon_{F}\rightarrow\overline{\epsilon}_{F}=\epsilon_{F}-\text{Re}\left[\Sigma_{4}\right]$. Expansion of function~\eqref{eq:self-energy averaging} in terms of $U_{0}F$ up to the second order gives the self-consistent Born approximation $\Sigma=\left(U_{0}^{2}/12\right)F$, used in the works~\cite{Liu2009,Groth,Guo}. Next, to obtain analytical expressions, we simplify the problem and set $\hat{G}_{\mathbf{k}}\rightarrow\hat{G}_{\mathbf{k}}^{0}$, where $\hat {G}_{\mathbf{k}}^{0}=\left(\epsilon_{F}+i\delta-H_{\mathbf{k}}^{0}\right)^{-1}$ . Then, using the same considerations as in the works~\cite{Liu2009,Groth,Guo} to calculate $\sum_{BZ}\hat{G}_{\mathbf{k}}^{0}$, we obtain~\footnote{The Fermi energy is localized inside the gap, therefore the imaginary part $F$, which is the density of bulk states of a pure system, is zero at zero temperature.} :
\begin{equation}
\text{Re}\left[F\right]\simeq-\frac{1}{2\pi}\frac{t\Gamma_{0}+\gamma}{t^{2}-\gamma^{2}}=F_{0}\Gamma_{0}+F_{1}.
\end{equation}

The problem of finding the quantities $\Sigma_{\mu}$ can be simplified if we first establish their form in the expression~\eqref{eq:self-energy}, and then carry out averaging over impurities. As a result we obtain
\begin{multline}
\Sigma_{0}=2\pi t+\frac{4\pi^{2}\gamma t}{U_{0}}\ln\left|\frac{\left(U_{0}-4\pi\gamma\right)^{2}-16\pi^{2}t^{2}}{\left(U_{0}+4\pi\gamma\right)^{2}-16\pi^{2}t^{2}}\right|\\
+\frac{2\pi^{2}\left(\gamma^{2}+t^{2}\right)}{U_{0}}\ln\left|\frac{\left(U_{0}-4\pi t\right)^{2}-16\pi^{2}\gamma^{2}}{\left(U_{0}+4\pi t\right)^{2}-16\pi^{2}\gamma^{2}}\right|,
\label{eq: SelfE0}
\end{multline}
where $\Sigma_{4}$ is defined  as
\begin{equation}
\Sigma_{4}=-\Sigma_{0}\left(\gamma\rightleftharpoons t\right)\label{eq:SelfE4}.
\end{equation}
One can obtain the same expressions if  one finds $\text{tr}\left(\Gamma_{\mu}\Sigma\right)$ from the formula~\eqref{eq:self-energy averaging}. Note that, in contrast to the Born approximation used in the works~\cite{Liu2009,Groth,Guo}, the expression~\eqref{eq: SelfE0},~\eqref{eq:SelfE4} for the self-energy part does not contain the singularity at $t=\gamma$. This is due to the fact that we summed up the entire single impurity diagram series (see Fig.~\ref{fig:diagram1}).

The main result of this paper, as will be seen below, is based on the dependence $\overline{m}\left(U_{0}\right)$. In this case, the condition $\overline{m}\leqslant0$ should be satisfied for edge states to exist. This condition can be obtained either in a conventional insulator ($m>0$) in the Anderson phase ($U_{0}^{2}>24\pi m\left(t^{2}-\gamma^{ 2}\right)/t$), or in a topological insulator ($m<0$) with $t<\gamma$. Further, we proceed from the dependence $\overline{m}\left(U_{0}\right)$, assuming that this condition is satisfied.

The Burkov-Balents model we use to study a multilayer system is based on the inclusion of tunnelling amplitudes between the edge modes of TI films. These edge modes represent zero-energy solutions. Here we consider the effect of disorder on the spatial distribution of these modes. We start from the Hamiltonian $H_{TI}^{0}$. It is necessary to solve a boundary value problem within open boundaries condition. This problem has been considered by many researchers (see, for example,~\cite{Zhou,Linder,Liu}) in the continuum approximation, when $d_{0}\approx\chi-6t+t\sum_{i}k_{i} ^{2}=m+tk^{2}$, $d_{i}\approx-2\lambda k_{i}=\upsilon k_{i}$, $d_{4}\approx0$. Then, the problem of finding edge modes is reduced to solving the wave equation with a one-dimensional Dirac Hamiltonian along an axis perpendicular to the edge. This approximation allows us to see all the basic properties of edge states. The qualitative picture of the behavior of edge states is as follows: at $m<0$ there are edge states, but at $m>0$ the edge states disappear, and at $m=0$ these edge states become bulk and the system is in the state of a Dirac semimetal. It is important that edge modes have a finite penetration depth, the latter being a function of the topological mass. This fact is of key importance for our work. The dependence of the penetration depth on the topological mass can be obtained from the solution of the boundary value problem~\cite{Zhou,Linder,Liu}: $\xi=\hbar\upsilon/\left|m\right|$. From this it is clear that $\xi\rightarrow\infty$ for $m\rightarrow0$, i.e. edge modes permeate the entire system, becoming bulk states. In the presence of non-magnetic disorder, in the Born approximation, taking into account the above results for the penetration depth, we obtain
\begin{figure}[t]
\includegraphics[width = 1\linewidth]{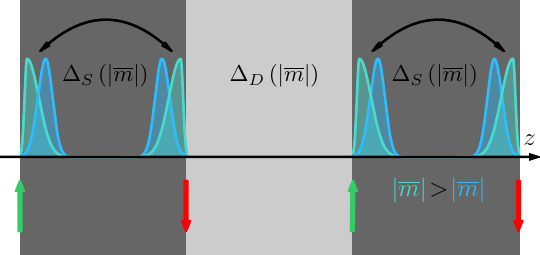}
\caption{Schematic representation of the dependence of the spatial distribution of edge modes on the topological mass value. A smaller value of $\left|\overline{m}\right|$ corresponds to a greater depth penetration.}
\label{fig:penetration dept}
\end{figure}

\begin{figure}[b]
\includegraphics[width = 1\linewidth]{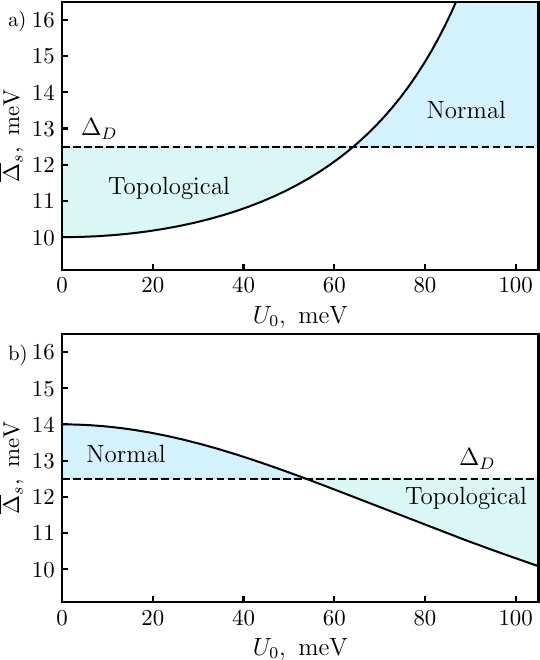}
\caption{Dependence of the value $\overline{\Delta}_{S}$ on $U_{0}$. The upper panel shows the case where $m<0$ and $t<\gamma$ and an increase in $U_{0}$ leads to an increase in $\overline{\Delta}_{S}$. It can be seen that $\overline{\Delta}_{S}=\Delta_{D}$ for some value of $U_{0}$. This leads to a transition from the topological phase to the normal. This transition can be understood if we use the formula for the $Z_{2}$ topological invariant from the work~\cite{FuKane}, rewritten as $(-1)^{\nu}=\text{sgn}\left(\overline {\Delta}_{S}-\Delta_{D}\right)$. The lower demonstrates the case when films in a multilayer structure are Anderson insulators with $m>0$ and $t>\gamma$, and an increase in the value of $U_{0}$ leads to an increase in the topological mass and a decrease in $\overline{\Delta }_{S}$.}
\label{fig:transition without Zeem}
\end{figure}
\begin{equation}
\overline{\xi}=\frac{\hbar\upsilon}{\left|m+\text{Re}\Sigma_{0}\right|},
\end{equation}
where $\text{Re}\Sigma_{0}$ is defined by Eq.~\eqref{eq: SelfE0}, while within Born approximation $\text{Re}\left[\Sigma_{0}\right]=-U_{0}^{2}t/24\pi(t^{2}-\gamma^{2})$~\cite{Liu,Groth}. Thus, the penetration depth of edge states is a function of disorder. This dependence is shown schematically in Fig.~\ref{fig:penetration dept}, for the case $m<0$ and $t<\gamma$. Our main idea is that non-magnetic disorder thus leads to a renormalization of tunnelling amplitudes between edge modes through a change in the penetration depth of these states. Two main mechanisms for such renormalization can be noted. Firstly, increasing the penetration depth leads to a decrease in the effective distance between edge states inside the TI film. This, accordingly, leads to an increase in the parameter $\Delta_{S}$. This also can be understood if we take into account that $\Delta_{S}\sim\int d^{3}\mathbf{r}\psi_{edge1}^{*}\left(\mathbf{r}\right)\psi_{edge2}\left(\mathbf{r}+L\right)$. Secondly, an increase in the overlap of edge modes as they approach each other further increases the amplitude of tunneling between them due to the effects of violation of orthogonality $\Delta_{S}\rightarrow\Delta_{S}/\left(1-S^{2}\right)$ ($S$ is the modulus of the overlap integral)~\cite{Harrison1,Harrison2}. Strictly speaking, renormalization of the penetration depth also affect $\Delta_{D}$. As can be seen from Fig.~\ref{fig:penetration dept}, this parameter decreases as $\overline{\xi}$ increases. However, this effect is of a higher order of smallness than the change in $\Delta_{S}$. For simplicity, we neglect it here. So, the Burkov-Balents model Hamiltonian in the presence of non-magnetic disorder can be rewritten as follows
\begin{equation}
\mathcal{\overline{H}}_{\mathbf{k}}=\left(\begin{array}{cc}
\upsilon_{F}\left(\hat{\bm{z}}\times\bm{\sigma}\right)\cdot\mathbf{k}_{\perp} & \overline{\Delta}_{S}+\Delta_{D}e^{ik_{z}d}\\
\overline{\Delta}_{S}+\Delta_{D}e^{-ik_{z}d} & -\upsilon_{F}\left(\hat{\bm{z}}\times\bm{\sigma}\right)\cdot\mathbf{k}_{\perp}
\end{array}\right).\label{B_B_hamiltonian_with_disorder}
\end{equation}
\begin{figure}[t]
\includegraphics[width = 1\linewidth]{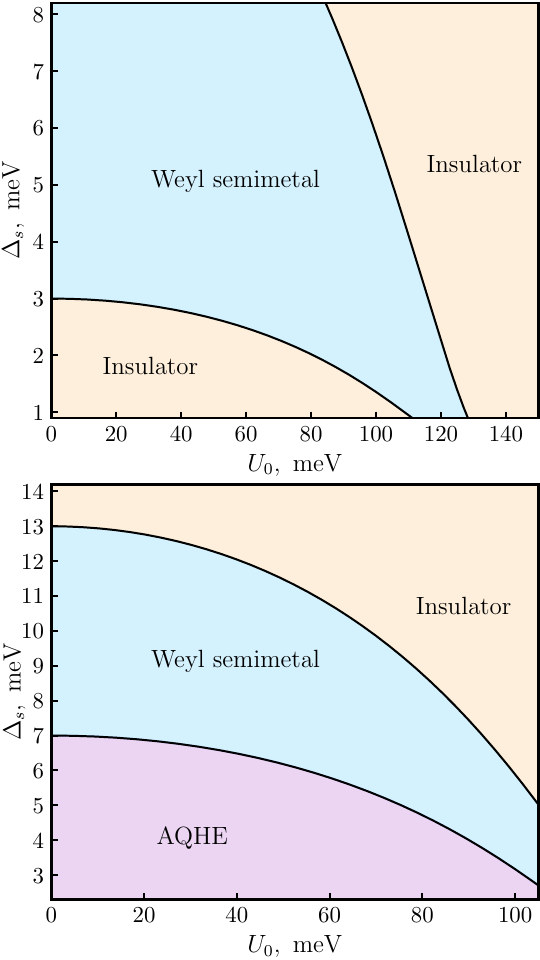}
\caption{Phase diagrams for the parameters $\Delta_{S}$ and $U_{0}$. The upper panel shows the situation $\Delta_{Z}<\Delta_{D}$, when only the phases of an normal insulator and a Weyl semimetal can be realized. The lower panel demonstrates the situation when $\Delta_{Z}>\Delta_{D}$, when in addition to the indicated phases, the implementation of the AQHE phase is possible. From the figures it is clear that disorder can be the cause of transitions between the mentioned phases.}
\label{fig:transition_with_Zeem}
\end{figure}

Let us get now the dependence of $\overline{\Delta}_{S}$ on the parameter $\overline{\xi}$. We proceed from the most general and simple consideration. For $\left| \overline{m}\right| \rightarrow\infty$, edge modes are localized directly near the edge, i.e. $\overline{\xi}\rightarrow 0$. 
In this case $\overline{\Delta}_{S}\left(\overline{\xi}=0\right)=\Delta_{S}=t_{S}e^{-\alpha L}$, where $L$ is the distance between edge modes, which coincides with the thickness of the TI film, $t_{S},\alpha$ are some constants characterizing tunneling (see~\cite{Miller}). For finite values of $m$, the effective distance between edge modes decreases to $L-2\overline{\xi}$. Thus, $\overline{\Delta}_{S}=\Delta_{S}e^{2\alpha\overline{\xi}}$, and $\overline{\xi}\leqslant L/2$. We use this dependence further. In Fig.~\ref{fig:transition without Zeem}, we show the phase diagrams of multilayer TI for the case of $m<0$ and $t<\gamma$. We assume that $\alpha=10^{-5}\text{cm}^{-1}$. It is clear from the figure that non-magnetic disorder can induce a phase transition from the topological phase to the trivial phase and vice versa, depending on the sign of the ordered topological mass and the relationships between the parameters $t,\gamma$.

Let us now move to the brief discussion of the influence of Zeeman splitting on the effects presented above. In this case, the Hamiltonian can be rewritten in the following form

\begin{equation}
\mathcal{\overline{H}}_{\mathbf{k}}\rightarrow\mathcal{\overline{H}}_{\mathbf{k}}+\tau^{0}\sigma_{z}\Delta_{Z},\label{eqn:hami_zeem}
\end{equation}
where $\Delta_{Z}$ is the value of Zeeman splitting. Zeeman splitting leads to the appearance of Weyl points, as well as to the transition to the AQHE phase (see,~\cite{Burkov}). The spectrum of such a Hamiltonian, accordingly, has the form
\begin{equation}
\varepsilon_{k\pm}^{2}=\upsilon_{F}^{2}\left(k_{x}^{2}+k_{y}^{2}\right)+\left[\Delta_{Z}\pm\overline{\Delta}\left(k_{z}\right)\right]^{2}.
\end{equation}
Here we once again emphasize that non-magnetic disorder inside the TI layers leads to a renormalization of the topological mass in them and, accordingly, the tunnelling parameters, and the Zeeman splitting $\Delta_{Z}$ is created, for example, due to the proximity effect of magnetic impurities in the dielectric layers (see, for example, ~\cite{Liu}). Thus, magnetic impurities do not affect the topological mass. From such a Hamiltonian one can obtain the condition for the existence of Weyl points~\cite{Burkov}. This condition has the form
\begin{equation}
\left(\overline{\Delta}_{S}-\Delta_{D}\right)^{2}<\Delta_{Z}^{2}<\left(\overline{\Delta}_{S}+\Delta_{D}\right)^{2}.
\end{equation}
It is clear that the fulfillment of this condition depends on disorder inside TI layers. On the upper panel in Fig.~\ref{fig:transition_with_Zeem}, one can see the phase diagram as a function of disorder $U_{0}$. It is clear from the figure that non-magnetic disorder can be the cause of a phase transition from the state of a normal insulator to the Weyl phase and vice versa. The main conclusion that follows from this is that uncontrolled disorder in real TI multilayer structures can be the reason for the implementation of different phases in them with the same composition and structure. This is especially important for the values of tunnelling parameters corresponding to the phase boundaries in this diagram. In the vicinity of such boundaries, even small fluctuations of these parameters can lead to interesting consequences. One type of such fluctuations is discussed in the next section.

\begin{figure}[t]
\includegraphics[width = 1\linewidth]{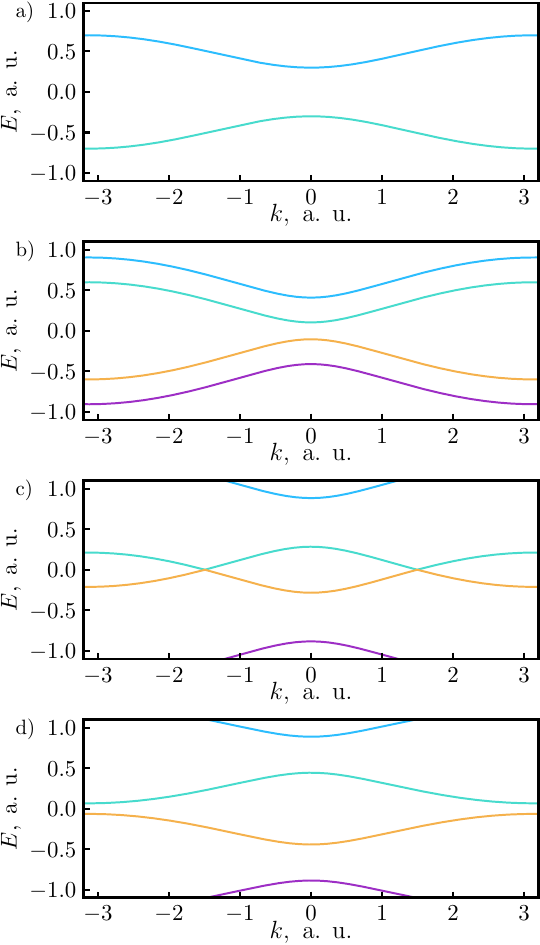}
\caption{Spectrum of Hamiltonian~\eqref{eqn:hami_zeem} for different parameter values: a) $\Delta_{Z}=0$ - an ordinary insulator, b) $\left|\Delta_{Z}\right|<\left|\overline{\Delta}_ {S}-\Delta_{D}\right|$ - ordinary insulator, c) $\left|\overline{\Delta}_{S}-\Delta_{D}\right|<\left|\Delta_{Z }\right|<\left|\overline{\Delta}_{S}+\Delta_{D}\right|$ - Weyl semimetal phase, d) $\left|\Delta_{Z}\right|>\left |\overline{\Delta}_{S}+\Delta_{D}\right|$, inverted spectrum - AQHE state.}
\label{fig:spectrum_with_Zeem} 
\end{figure}

Finally, let us dwell briefly on transitions to the AQHE state, which arises separately from the Weyl phase and the phases of the topological and normal insulators. This phase occurs at $\Delta_{Z}^{2}>\left(\overline{\Delta}_{S}+\Delta_{D}\right)^{2}$. The lower panel of Fig.~\ref{fig:transition_with_Zeem} shows the phase diagram for the values of such tunnelling parameters and Zeeman splitting at which disorder-induced transitions between different phases are possible. It can be seen that the Weyl semimetal phase appears as an intermediate phase between the normal insulator ($\Delta_{Z}^{2}<\left(\overline{\Delta}_{S}-\Delta_{D}\right)^{2 }$) and the state of AQHE. It should be noted that the anomalous Hall effect appears already in the Weyl phase. In this case, the Hall conductivity is proportional to the distance between the Weyl points. In turn, the latter depends on the relationship between Zeeman splitting and tunnelling parameters. At the same time, at $\Delta_{Z}^{2}>\left(\overline{\Delta}_{S}+\Delta_{D}\right)^{2}$ a quantized Hall state (Hall conductivity plateau) arises. From the spectra in Fig.~\ref{fig:spectrum_with_Zeem} it can be understood that in the Weyl phase the anomalous Hall conductivity arises due to the presence of Weyl points, and at $\Delta_{Z}^{2}>\left(\overline{\Delta}_ {S}+\Delta_{D}\right)^{2}$ the spectrum becomes inverted, which leads to a non-zero Chern number and a Hall plateau. It should be noted that disorder-induced transitions between the phases of a Weyl semimetal, AQHE, and normal insulator in a double quantum well were predicted in~\cite{Krishtopenko}. The nature of such transitions in a multilayer structure, studied in our paper, is absolutely new. The effects we predict are based on changes in the interlayer tunneling parameters due to renormalization of the penetration depth of edge modes.

\section{Off-diagonal disorder in Multilayer Topological Insulator}
\label{sec:IV}

In this section, we consider the situation when the characteristics of impurities (potential, concentration) fluctuate randomly from layer to layer. In this case, we are dealing with the Burkov-Balents model with fluctuating tunnelling parameters. The type of disorder when the fluctuating quantity is not the site energy (the diagonal parameter of the Hamiltonian) but the inter-sites hopping is called as the non-diagonal disorder~\cite{Raghavan,Harris,Ziman}. Historically, the inclusion of off-diagonal disorder in the theory of disordered electron (and phonon) systems was apparently done in~\cite{SHIBA,Blackman,Moorjani}. In these works, an approximation of the coherence potential with off-diagonal disorder was developed for binary transition metal alloys. The main reason why the off-diagonal disorder occurs in such systems is that the width of the $d$-level increases the higher the $d$-level. Therefore, to realistically describe the electronic structure of transition metal alloys, it may be important to take into account the difference in the hopping integral, in addition to the difference in $d$-level energies (e.g., a $3d$-$3d$ hopping is different from a $3d$-$4d$ hopping). There are also alloys of other metals in which off-diagonal disorder cannot be neglected. In our system, the off-diagonal disorder is caused by the difference in the penetration depths of edge modes in different layers, caused by the diagonal disorder inside TI layers. 

It is also necessary to note the following important feature of multilayer TI. The interlayer transport occurs due to tunneling between chiral edge modes. These modes themselves are protected against scattering by nonmagnetic impurities. Thus, we are dealing with a system in which delocalized states in the plane (edge modes) do not sense the presence of disorder. Thus, disorder only affects interlayer states.

\begin{figure}[t]
\includegraphics[width = 0.99\linewidth]{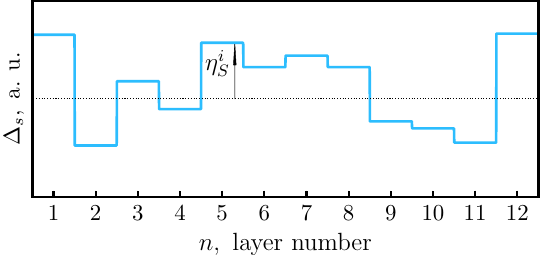}
\caption{Schematic representation of off-diagonal disorder: tunnel parameter $\Delta_S$ as a random function of layer number.}
\label{fig:fluctuations}
\end{figure}

Let us formulate the problem for the case when the off-diagonal disorder presents in the system. In the most general form, the Burkov-Balents Hamiltonian with fluctuating tunnelling parameters can be written as
\begin{multline}
\mathcal{H}=\sum_{\mathbf{k}_{\perp},i}c_{\mathbf{k}_{\perp}i}^{\dagger}\left[\tau^{z}\upsilon_{F}\left(\hat{\bm{z}}\times\bm{\sigma}\right)\cdot\mathbf{k}_{\perp}+\Delta_{S}^{i}\tau^{x}\right]c_{\mathbf{k}_{\perp}i}\\
\!\!+\frac{1}{2}\sum_{\mathbf{k}_{\perp},i,j}c_{\mathbf{k}_{\perp}i}^{\dagger}\!\left[\Delta_{+}^{i}\tau^{+}\delta_{j,i+1}+\frac{1}{2}\Delta_{-}^{i}\tau^{-}\delta_{j,i-1}\right]c_{\mathbf{k}_{\perp}j}\label{off-diagonal Hamiltonian},
\end{multline}
where the quantities $\Delta_{S}^{i},\Delta_{\pm}^{i}$ are random functions of the number of the layer $i$ of the TI. Next we can write $\Delta_{S}^{i}=\Delta_{S}+\eta_{S}^{i}$ and $\Delta_{\pm}^{i}=\Delta_{D}+ \eta_{\pm}^{i}$, where $\Delta_{S},\Delta_{D}$ are parts of the tunnelling parameters that are regular and identical for all layers, and $\eta{}_{S}^{i },\eta{}_{\pm}^{i}$ are corrections (fluctuations) that are random functions of the layer number. However, as in the previous part of the article, we consider here a simpler case, when only $\Delta_{S}$ is the fluctuating quantity. In this case, the fluctuations are similar to the random potential on the layer, because this quantity is included with the factor $\delta_{ij}$, although $\Delta_{S}$ is the off-diagonal parameter and describes tunneling between edge modes, which is expressed by the presence of a factor in the form of the Pauli matrix $\tau^{x}$. Neglecting fluctuations in $\Delta_{D}$ simplifies the problem. However, such neglect does not lead to the loss of important features of the system. These fluctuations are shown schematically in Fig.~\ref{fig:fluctuations}. Let the quantity $\eta{}_{S}^{i}$ take values from the interval $\left(-\eta{}_{0}/2,\eta{}_{0}/2\right)$ with uniform distribution. From simple topological considerations, three modes can be distinguished: 1) $\overline{\Delta}_{S}-\eta{}_{0}/2>\Delta_{D}$, 2) $\overline{\Delta} _{S}+\eta{}_{0}/2<\Delta_{D}$, 3) $\overline{\Delta}_{S}+\eta{}_{0}/2>\Delta_ {D}>\overline{\Delta}_{S}-\eta{}_{0}/2$. However, these modes are not clearly manifested for all phases. Below we calculate the density of states for different phases as a function of disorder. We expect that radically new phenomena are possible in such a system, for example, the induced by fluctuations appearance of delocalized states inside the band gap, up to the total collapse of the gap, etc. Such effects for a one-dimensional semiconductor system with a fluctuating gap were predicted in the works~\cite{Ovchinnikov,Gredeskul}. Below we see that something similar occurs in our system.

We assume that fluctuations $\eta{}_{S}^{i}$ are independent of each other, i.e. $\left\langle \eta{}_{S}^{i}\eta{}_{S}^{j}\right\rangle =D\delta_{ij}$. Thus, formally we are dealing with an Anderson-type model, in which the fluctuating \guillemotleft potential\guillemotright{} has the form $\eta{}_{S}^{i}\tau^{x}$.

The Appendix~\ref{app:1} shows the calculation of the self-energy part in the simplest Born approximation without taking into account Zeeman splitting. Within this approximation, it is already clear that off-diagonal disorder gives an additional renormalization of the Hamiltonian parameters. As we can see, such renormalization is proportional to the squared fluctuation of the tunnelling parameter $\eta{}_{0}^{2}$. As noted above, fluctuations of the tunnelling parameter can lead to the so-called effects of off-diagonal Anderson localization/delocalization~\cite{Raghavan,Harris,Ziman}. Therefore, below we dwell in more detail on the study of the influence of this type of disorder.

For a more detailed study of the effects of off-diagonal disorder and to identify the possibility of the emergence of new states induced by it, we use the method of Green function expansion in terms of the hopping integral, developed in the works~\cite{Toyozawa,Matsubara} and presented in the review~\cite{J.M.Ziman}. It was shown in~\cite{Leath} that this method, with self-consistent consideration of the expansion parameters, is completely equivalent to the expansion in terms of the impurity potential. In our case, the disorder is off-diagonal, so this method seems more convenient. In the case of off-diagonal disorder in binary alloys, this method was generalized in the works~\cite{SHIBA,Blackman} (see also~\cite{Moorjani,Yonezawa}). Let us rewrite the Hamiltonian~\eqref{off-diagonal Hamiltonian} into the following form
\begin{equation}
\mathcal{H}=\sum_{\mathbf{k}_{\perp},i,j}c_{\mathbf{k}_{\perp}i}^{\dagger}\left[T_{ij}\left(\mathbf{k}_{\perp}\right)+V_{i}\delta_{i,j}\right]c_{\mathbf{k}_{\perp}j},
\end{equation}
where
\begin{align}
T_{ij}\left(\mathbf{k}_{\perp}\right)&=\left(\tau^{z}\upsilon_{F}\left(\hat{\bm{z}}\times\bm{\sigma}\right)\cdot\mathbf{k}_{\perp}+\Delta_{S}\tau^{x}\right)\delta_{i,j}\nonumber\\
&\qquad+\frac{1}{2}\Delta_{D}\tau^{+}\delta_{j,i+1}+\frac{1}{2}\Delta_{D}\tau^{-}\delta_{j,i-1},\\
V_{i}&=\eta{}_{S}^{i}\tau^{x}.
\end{align}

Following~\cite{Toyozawa,Matsubara,J.M.Ziman}, we expand the Green function in the site representation in terms of the interlayer hopping integral $t_{ij}=\frac{1}{2}\Delta_{D}\left(\tau^{+ }\delta_{j,i+1}+\tau^{-}\delta_{j,i-1}\right)$ (method of expansion in localized states):
\begin{multline}
G_{ij}\left(\mathbf{k}_{\perp},\epsilon\right)=S_{i}\left(\mathbf{k}_{\perp}\right)\delta_{ij}+S_{i}\left(\mathbf{k}_{\perp}\right)t_{ij}S_{j}\left(\mathbf{k}_{\perp}\right)\\
+\sum_{j'}S_{i}t_{ij'}S_{j'}t_{j'j}S_{j}+\dots ,
\end{multline}
where
\begin{equation}
S_{i}\!\left(\mathbf{k}_{\perp}\!\right)\!=\!\left(\!\epsilon\!-\tau^{z}\upsilon_{F}\left(\hat{\bm{z}}\times\bm{\sigma}\right)\cdot\mathbf{k}_{\perp}\!-\!\Delta_{S}\tau^{x}-V_{i}\right)^{-1}.
\end{equation}
Accordingly, for the Fourier transform we obtain
\begin{multline}
G_{k_{z}k_{z}'}\left(\mathbf{k}_{\perp},\epsilon\right)=\sigma_{k_{z}k_{z}'}+\sum_{k_{z}''}\sigma_{k_{z}k_{z}''}t_{k_{z}''}\sigma_{k_{z}''k_{z}'}\\
+\sum_{k_{z}'',k_{z}'''}\sigma_{k_{z}k_{z}''}t_{k_{z}''}\sigma_{k_{z}''k_{z}'''}t_{k_{z}'''}\sigma_{k_{z}'''k_{z}'}+...\label{Green_func_equation}
\end{multline}
where $\sigma_{k_{z}k_{z}'}=\frac{1}{N}\sum_{j}S_{j}\exp\left[i\left(k_{z}-k_{z }'\right)z_{j}\right]$ and $t_{k_{z}}=\Delta_{D}\left(\tau^{x}\cos\left(k_{z}a\right) -\tau^{y}\sin\left(k_{z}a\right)\right)$. Next, it is necessary to average such a Green function over disorder. If the value of $\eta{}_{S}^{i}$ were the same in each layer, then the matrix $\sigma_{k_{z}k_{z}'}$ would be diagonal in momenta, which would allow us factor the terms and sum the entire series. However, $\eta{}_{S}^{i}$, and therefore $S_{i}\left(\mathbf{k}_{\perp}\right)$ are stochastic quantities, the ensemble average of the products which can not be equated to the products of their averages. However, first of all, we use an approximation in which we replace all factors $\sigma_{k_{z}k_{z}'}$ with average ones according to the following formula (below we use a more accurate approach):
\begin{equation}
\left\langle \sigma_{k_{z}k_{z}'}\right\rangle =\left\langle S_{j}\right\rangle \delta_{k_{z}k_{z}'},
\end{equation}
which leads to
\begin{equation}
\left\langle G_{k_{z}}^{\left(0\right)}\left(\mathbf{k}_{\perp},\epsilon\right)\right\rangle =\left(\left\langle S_{j}\right\rangle ^{-1}-t_{k_{z}}\right)^{-1}\label{simplest approx}.
\end{equation}

It should be noted here that when averaging this kind of series of expansions of Green functions, it is important to correctly split the Hamiltonian into the unperturbed part and the perturbation. A rougher approximation can be obtained if we use the expansion
\begin{equation}
\left\langle S_{j}\right\rangle^{-1}\! \approx \!
\left\langle\! \left(\!\epsilon\!-\!V_{i}\!\right)^{-1}\!\right\rangle ^{-1}\!\!-\tau^{z}\upsilon_{F}\left(\hat{\bm{z}}\times\bm{\sigma}\right)\cdot\mathbf{k}_{\perp}-\Delta_{S}\tau^{x},
\end{equation}
which corresponds to the expansion of the Green function in terms of the complete jump integral $T_{ij}$. However, the applicability of such an expansion requires the smallness of all parameters included in $T_{ij}$, which is why it is rougher. 

As we noted above, we will use Anderson approach when the random variable $\eta{}_{S}^{i}$ is distributed uniformly in the interval $\left(-\eta{}_{0}/2,\eta {}_{0}/2\right)$. Then
\begin{equation}
\left\langle S_{j}\right\rangle =\frac{1}{\eta{}_{0}}\int_{-\eta{}_{0}/2}^{\eta{}_{0}/2}S_{i}\left(\mathbf{k}_{\perp},\eta{}_{S}^{j}\right)d\eta{}_{S}^{j},
\end{equation}
where it is assumed that $\epsilon=\epsilon+i0$. The value $\text{Re}\left[\left\langle S_{j}\right\rangle ^{-1}\right]$ leads to renormalization of the shape of the band structure, and the value $\text{Im}\left\langle S_{j}\right \rangle ^{-1}$ is the frequency of scattering by fluctuations of the tunnelling parameter. This quantity goes to zero outside the fluctuation range of energy values. Thus
\begin{gather}
\left\langle G_{k_{z}}\left(\mathbf{k}_{\perp},\epsilon\right)\right\rangle =G_{k_{z}}^{0}\left(\left\langle S_{j}\left(\mathbf{k}_{\perp}\right)\right\rangle ^{-1}\right).
\end{gather}
We should note here that $\lim_{\eta{}_{0}\rightarrow0}\left\langle S_{j}\right\rangle =\left(\epsilon-\tau^{z}\upsilon_{F}\left (\hat{\bm{z}}\times\bm{\sigma}\right)\cdot\mathbf{k}_{\perp}-\Delta_{S}\tau^{x}\right)^{ -1}$, which leads to an expression for the Green functions of an ordered system. To calculate the density of states we use the standard expression
\begin{gather}
\rho\left(\epsilon\right)=-\pi^{-1}\text{Im}\sum_{\mathbf{k}_{\perp},k_{z}}\text{tr}\left(\left\langle G_{k_{z}}\left(\mathbf{k}_{\perp},\epsilon+i0\right)\right\rangle \right).
\end{gather}
\begin{figure}[t]
\includegraphics[width = 1\linewidth]{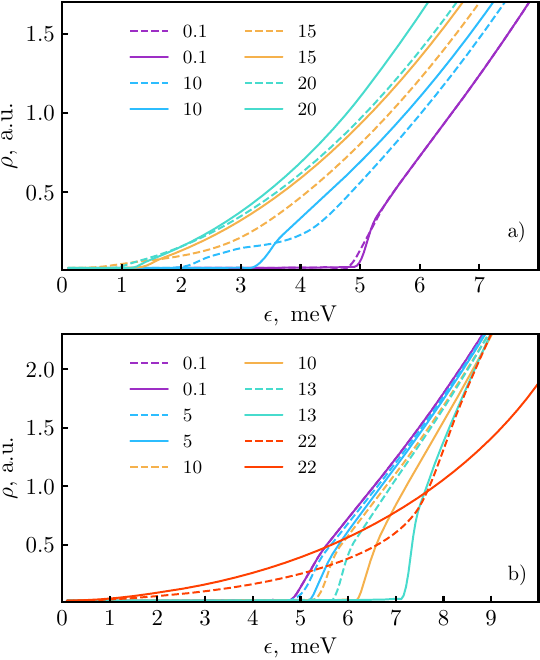}
\caption{Density of states as a function of energy for two phases in the absence of Zeeman splitting ($\Delta_{Z}=0$). The upper panel demonstrates normal insulator phase with $\Delta_{S}=25\text{meV}$ and $\Delta_{D}=20\text{meV}$. The band gap regions corresponding to different density of states curves are shaded in different colors. An increase in $\eta_{0}$ leads to a decrease in the band gap and the appearance of delocalized states. The lower panel corresponds  to topological insulator phase with $\Delta_{S}=20\text{meV}$ and $\Delta_{D}=25\text{meV}$. An increase in $\eta_{0}$ first leads to an increase in the band gap (to approximately $\eta_{0}=13\text{meV}$), and then the band gap quickly collapses, also leading to delocalized states. Thus, the topological phase is stable at small values of $\eta_{0}$. Only positive energies are shown, because the curves are symmetrical about the vertical axis.}
\label{fog:DOS_without_Zeem}
\end{figure}

In Fig.~\ref{fog:DOS_without_Zeem}, we show the density of states for the cases where the ideal system is a normal insulator (top panel) and a topological insulator (lower  panel). The solid lines correspond to the approximation~\eqref{simplest approx}, and the dotted lines are obtained in a more rigorous approach, which is presented below. It is interesting that the phases of normal and topological insulators differ from each other in their response to fluctuations of the tunnel parameter. Namely, the width of the gap of a normal insulator decreases with increasing value $\eta_{0}$ until it is completely closed. On the other hand, the gap width of a topological insulator first increases (up to $\eta_{0}=13 \text{meV}$ in the figure), and then collapses quite quickly with increasing value $\eta_{0}$. The general conclusion that we can draw is that fluctuations in the value of $\eta{}_{S}^{i}$ lead to the emergence of delocalized states inside the gap, but in the topological phase such states arise only at large values of fluctuations. Thus, we obtain an interesting effect, which in essence can be called Anderson delocalization.

\begin{figure}[t]
\includegraphics[width = 1\linewidth]{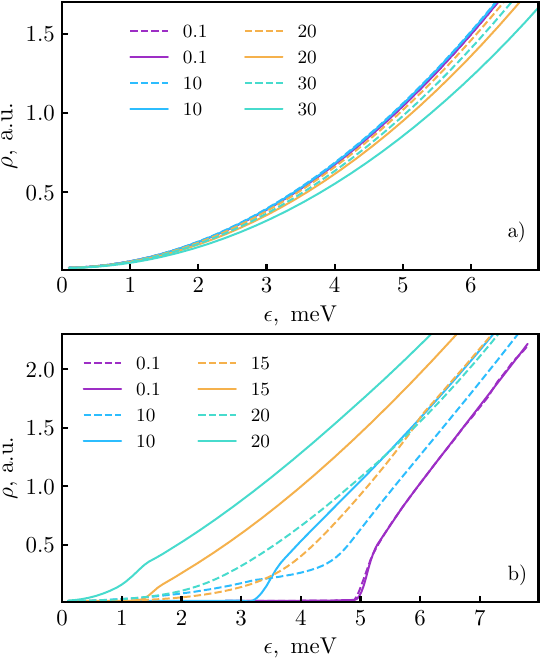}
\caption{Density of states as a function of energy in the presence of Zeeman splitting ($\Delta_{Z}\protect\neq0$) at $\Delta_{S}=25\text{meV}$ and $\Delta_{D}=20\text{meV }$. The upper panel demonstrates Weyl semimetal phase ($\left|\overline{\Delta}_{S}-\Delta_{D}\right|<\Delta_{Z}=20\text{meV}<\left|\overline{\Delta}_{S}+\Delta_{D}\right|$). The density of states changes very little with increasing $\eta_{0}$. Thus, the Weyl point is stable to non-diagonal disorder. The lower panel shows AQHE phase ($\Delta_{Z}=50\text{meV}>\left|\overline{\Delta}_{S}+\Delta_{D}\right|$). It can be seen that as $\eta_{0}$ increases, the band gap only increases, which indicates the stability of the AQHE phase relative to off-diagonal disorder.}
\label{fig:DOS with Zeem}
\end{figure}

Let us now   include the Zeeman term in the Hamiltonian and study the stability of the Weyl phase and the AQHE phase  by the  following inclusion $S_{j}^{-1}\to S_{j}^{-1}-\tau^{0}\sigma_{z}\Delta_{Z}$. In Fig.~\ref{fig:DOS with Zeem}, we  present the density of states in the Weyl phase (upper panel) and the AQHE phase (lower panel) for different values of $\eta_{0}$. The simplest approximation is depicted by solid lines. As can be seen from the figure, the Weyl phase is stable with respect to off-diagonal disorder even at large values of $\eta_{0}$. For the AQHE phase, an increase in the value of $\eta_{0}$ leads to the appearance of delocalized states, as in the cases of normal and topological insulators.

\begin{figure}[t]
\includegraphics[width = 1\linewidth]{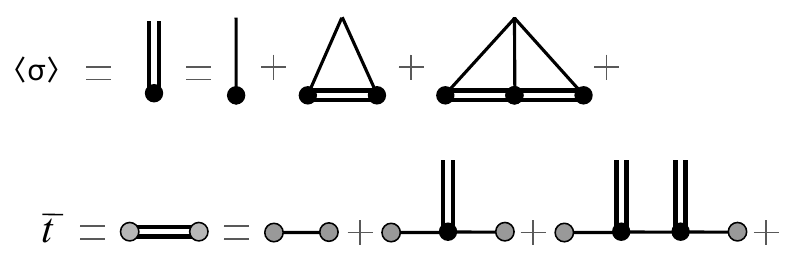}
\caption{Diagrammatic expansions for the quantities $\left\langle \sigma\right\rangle $ and $\overline{t}$. Gray circles in the lower diagrams indicate identical indices. Intermediate indices can be any, but differ from the edge indices (gray). This requirement simply corresponds to the exclusion of repeated counting of the same charts.}
\label{fig:Diagram series}
\end{figure}

Above we used the simplest approximation, replacing all functions $\left\langle \sigma_{k_{z}k_{z}^{\left(1\right)}}\dots \sigma_{k_{z}^{\left(n\right)}k_{z}'}\right\rangle$  during averaging \eqref{Green_func_equation} over disorder by products of averages $\left\langle S_{j}\right\rangle ^{n} \delta_{k_{z}k_{z}'}$. This is a rather rough approximation, similar in some sense to the mean field approximation (in the context of disorder physics, one can also make an analogy with the virtual crystal approximation). Here we refine our theory using the ideas of the generalized coherent potential approximation (see~\cite{J.M.Ziman,Matsubara,Blackman}). Let us introduce the renormalized locator $\left\langle \sigma\right\rangle $ and interactor $\overline{t}_{ij}$, as shown in Fig.~\ref{fig:Diagram series}. The renormalized locator $\left\langle \sigma\right\rangle $ contains partial summation over repeated indices (similar to how the scattering amplitude in the Born approximation is replaced by the scattering matrix)~\cite{Blackman}. This replacement takes into account that the series~\eqref{Green_func_equation} contains arbitrary powers of magnitude $S_{j}$, which, when averaging, do not reduce to products of averages $\left\langle S_{j}\right\rangle$. Next, we obtain the basic formulas for our more accurate theory.

To carry out partial summation over repeating indices, one can apply cumulative averaging (see~\cite{J.M.Ziman,Matsubara}). In our case, additional caution must be exercised due to the matrix nature of the quantities $S_{j}$ and $T_{ij}$. For cumulative averaging we have
\begin{equation}
\!\!\!\left\langle \sum_{k_{z}''k_{z}'''}\!\!\sigma_{k_{z}k_{z}''}\overline{t}_{k_{z}''}\sigma_{k_{z}''k_{z}'''}\overline{t}_{k_{z}'''}\sigma_{k_{z}'''k_{z}'}\!\!\right\rangle _{c}\!\!\!\!=\!\left\langle \! S_{m}\overline{t}S_{m}\overline{t}S_{m}\!\right\rangle \!\delta_{k_{z}k_{z}'},\!\!\!\!
\end{equation}
where we used the cumulant property $\left\langle S_{i}\overline{t}S_{j}\overline{t}S_{s}...\right\rangle _{c}\neq0$ only for $i =j=s=...$. Taking this into account, we write the renormalized locator as follows
\begin{multline}
\left\langle \sigma\right\rangle =\left\langle S_{m}\right\rangle +\left\langle S_{m}\overline{t}S_{m}\right\rangle \\
+\left\langle S_{m}\overline{t}S_{m}\overline{t}S_{m}\right\rangle +...=\left\langle \left(S_{m}^{-1}-\overline{t}\right)^{-1}\right\rangle.\label{locator} 
\end{multline}
Finally, the diagrammatic series in Fig.~\ref{fig:Diagram series} gives the following expression for the value of $\overline{t}$
\begin{equation}
\overline{t}\left(\mathbf{k}_{\perp}\right)=\left\langle \sigma\right\rangle ^{-1}\left(\left\langle G\right\rangle t\right)_{mm}=\left\langle \sigma\right\rangle ^{-1}\sum_{k_{z}}\left\langle \overline{G}_{\mathbf{k}}\right\rangle t_{k_z}\label{interactor}
\end{equation}
where for the Green function we have
\begin{equation}
\left\langle \overline{G}_{\mathbf{k}}\right\rangle =\left(\left\langle \sigma\right\rangle ^{-1}+\overline{t}\left(\mathbf{k}_{\perp}\right)-t_{k_{z}}\right)^{-1}\label{self_green_func}.
\end{equation}
In the absence of disorder $\left\langle \sigma\right\rangle =\left(S_{m}^{-1}-\overline{t}\left(\mathbf{k}_{\perp}\right)\right)^{-1}$ and we obtain an expression for the unperturbed Green function. Thus, to calculate the density of states, we use the formulas \eqref{locator}-\eqref{self_green_func}, and carry out averaging over a homogeneous ensemble, i.e. similar to Anderson model. The method described by the formulas \eqref{locator}-\eqref{self_green_func} is called the self-consistent locator method.

If we put $G=S_{m}$ in \eqref{interactor}, we obtain $\overline{t}=0$ and $\left\langle \sigma\right\rangle =\left\langle S_{m }\right\rangle $, which brings us to the simplest approximation we used above. To obtain the next approximation, we can put in \eqref{interactor} $G=G^{1}$, where $G^{1}$ is the Green function obtained in the simplest approximation, i.e. $G_{k}^{1}=\left(\left\langle S\right\rangle^{-1}-t_{\mathbf{k}}\right)^{-1}$. In Fig.~\ref{fog:DOS_without_Zeem} and~\ref{fig:DOS with Zeem} dotted lines show numerical calculations of the density of states using formulas~\eqref{locator}-\eqref{self_green_func}. As can be seen from the figures, in some cases the differences between the approaches are not small. But qualitatively the simplest approximation correctly describes the effect of gap collapse at large values of fluctuation of the tunnel parameter. It should be noted that when more precise formulas are used, the collapse of the gap increases. For example, for the phase of a normal insulator (see Fig.~\eqref{fog:DOS_without_Zeem} top) at $\eta_0=15$ meV in the zero approximation there is still a gap (about $1.2$ meV), but according to the dotted orange line the gap is narrower practically absent.

\section{Conclusion}

In this work, we investigated the effect of diagonal and off-diagonal non-magnetic disorder on different phases of a multilayer TI. We have shown that disorder can cause transitions between these phases. Off-diagonal disorder can induce states to arise inside the gap and even cause the band gap collapse. The appearance of delocalized states inside the bandgap can lead to a radical restructuring of the AQHE phase.

The appearance of delocalized states in the gapped phase can be studied experimentally using transport measurements. In particular, the temperature behavior of static interlayer conductivity is likely to be different for localized and delocalized states. In addition, information about the presence of such states caused by disorder can be obtained using measurements of thermoEMF and the Hall effect.

Finally, it is interesting to compare our results with the works of~\cite{Ovchinnikov,Gredeskul}. In the work~\cite{Ovchinnikov}, the density of states of a one-dimensional semiconductor system with a band gap, which is a random function of the coordinate, was studied. The authors found an exact solution, one of the consequences of which is the emergence of a singularity at the center of the gap $\Delta(x)=\Delta_{0}+\Delta'(x)$. According to~\cite{Ovchinnikov}, a singularity of the form $\left|\epsilon\ln^{3}\left|\epsilon\right|\right|^{-1}$ of the density of states at the center of the gap $\epsilon=0$ arises in infinitely long system with $\Delta_{0}=0$. In our case, the random variable is the tunnelling parameter $\Delta_{S}$. However, if one remembers that the band gap in our model is written as $\left|\overline{\Delta}_{S}-\Delta_{D}\right|$, then fluctuations of the parameter $\Delta_{S}$ can be perceived as fluctuations of the band gap. Thus, the analogy between our work and paper~\cite{Ovchinnikov,Gredeskul} becomes obvious. In fact, using a similarity transformation, the Hamiltonian~\eqref{B_B_hamiltonian_with_disorder} can be reduced to the form
\begin{gather}
\left(U^{-1}\mathcal{\overline{H}}_{\mathbf{k}}U\right)_{\mathbf{k}_{\perp}=0}=\left(\begin{array}{cc}\overline{\Delta}_{S}-\Delta_{D} & \Delta_{D}\nabla_{z}\\\Delta_{D}\nabla_{z} & -\left(\overline{\Delta}_{S}-\Delta_{D}\right)\end{array}\right),
\end{gather}
where we used the long-wavelength approximation and set $d=1$, $k_z\to -i\nabla_{z}$. This Hamiltonian completely coincides with what was studied in the work~\cite{Ovchinnikov}. The absence of a singularity of the density of states at the center of the band gap in our model is apparently due to the three-dimensionality of our problem and the fact that, unlike~\cite{Ovchinnikov}, we took into account the crystal structure along the $z$-axis, which leads to a limitation of the range of values of $k_{z}$. In addition, we mainly used perturbative methods, in contrast to~\cite{Ovchinnikov} where an exact solution was obtained. The study of off-diagonal disorder in multilayer TI using non-perturbative methods (for this, apparently, the method used in the work of~\cite{Bartosch} can be applied) and the study of the singular behavior of the density of states is a separate interesting problem.

\begin{acknowledgments}
The results of Chapters III and IV were obtained with the support of the Russian Science Foundation grant No. 22-72-00110. The study is partly supported by the Ministry of Science and Higher Education of the Russian Federation (Goszadaniye), project No. FSMG-2023-0011. 
\end{acknowledgments}

\appendix
\section{Appendix}\label{app:1}

Let us expand the operators $c_{\mathbf{k}_{\perp}i}^{\dagger},c_{\mathbf{k}_{\perp}i}$ into eigenfunctions of the unperturbed Hamiltonian
\begin{gather}
c_{\mathbf{k}_{\perp}i}=\sum_{k_{z}}c_{\mathbf{k}_{\perp}k_{z}}e^{-ik_{z}z_{i}}.
\end{gather}
Then
\begin{gather}
V=\sum_{\mathbf{k}_{\perp}k_{z}k_{z}'}c_{\mathbf{k}_{\perp}k_{z}'}^{\dagger}\left\langle k_{z}'\right|\eta\left|k_{z}\right\rangle c_{\mathbf{k}_{\perp}k_{z}},
\end{gather}
where $\left\langle k_{z}\right|\eta\left|k_{z}'\right\rangle =\sum_{i}e^{ik_{z}'z_{i}}\eta{} _{S}^{i}\tau^{x}e^{-ik_{z}z_{i}}$. If we introduce the Fourier transform of the quantity $\eta{}_{S}^{i}$ in the form $\eta{}_{S}^{i}=\sum_{q}\eta{}_{S}^{q}e^{-iqz_{i}}$,
then we obtain $\left\langle k_{z}\right|\eta\left|k_{z}'\right\rangle =\sum_{q}\eta{ }_{S}^{q}\tau^{x}\delta_{k_{z}',k_{z}+q}$. In this case
\begin{gather}
V=\sum_{\mathbf{k}_{\perp}k_{z}q}c_{\mathbf{k}_{\perp}k_{z}+q}^{\dagger}\left(\eta{}_{S}^{q}\tau^{x}\sigma_{0}\right)c_{\mathbf{k}_{\perp}k_{z}}.
\end{gather}
This shows that if $\eta_{S}^{i}=\eta_{S}$, then $\eta_{S}^{q}=\delta_{q0}$, which gives $V=\eta{ }_{S}\sum_{\mathbf{k}_{\perp}k_{z}q}c_{\mathbf{k}_{\perp}k_{z}}^{\dagger}\tau^{ x}c_{\mathbf{k}_{\perp}k_{z}}$, i.e. simply an addition in the form of a constant to the $\tau^{x}$-component of the zero Hamiltonian. Using perturbation theory with such a small parameter it is possible to obtain corrections of any order to the Hamiltonian. Thus, in addition to scattering on impurities, which leads to renormalization of tunnel parameters, scattering occurs on fluctuations of the impurity potential from layer to layer. Using the standard Matsubara formalism of Green's functions, for the second-order correction we obtain \footnote{For convenience, we omit the index $\mathbf{k}_{\perp}$ and the corresponding summation, which we will restore at the end.}
\begin{gather}
-\left\langle T\left[c_{k_{1}}\left(\tau_{1}\right)\otimes\overline{c}_{k_{2}}\left(\tau_{2}\right)V\left(\tau'\right)V\left(\tau''\right)\right]\right\rangle,
\end{gather}
where the tensor product acts in the matrix basis $4\times4$, in which the Hamiltonian is written, and the index D denotes . To obtain an analytical expression, we average over $\eta{}_{S}^{i}$, integrating over all $\eta{}_{S}^{i}$ with weight $1/\eta{} _{0}$
\begin{equation}
\left\langle \text{...}\right\rangle _{c}=\prod_{i}\int\limits_{-\eta{}_{0}/2}^{\eta{}_{0}/2}\frac{d\eta{}_{S}^{i}}{\eta{}_{0}}... \ .
\end{equation}
It is clear that in the first order in perturbation such averaging gives a zero correction to the Green function. In second order we have
\begin{multline}
\left\langle V^{2}\right\rangle _{c}\sim\sum_{qq'}\left\langle \eta{}_{S}^{q}\eta{}_{S}^{q'}\right\rangle _{c}f\left(q,q'\right)\\
=\sum_{ij}\left\langle \eta{}_{S}^{i}\eta{}_{S}^{j}\right\rangle _{c}\sum_{qq'}e^{iqz_{i}}e^{iq'z_{j}}f\left(q,q'\right)\\
=\frac{\eta{}_{0}^{2}}{12}\sum_{qq'}\delta_{q,-q'}f\left(q,q'\right)=\frac{\eta{}_{0}^{2}}{12}\sum_{q}f\left(q,-q\right).
\end{multline}
Then,  we have
\begin{multline}
-\frac{\eta{}_{0}^{2}}{12}\sum_{k_{z}'k_{z}''q}\left\langle T\left[c_{k_{1}\tau_{1}}\otimes\overline{c}_{k_{2}\tau_{2}}\overline{c}_{k_{z}'+q,\tau'}\left(\tau^{x}\sigma_{0}\right)c_{k_{z}'\tau'}\right.\right.\\
\times\left.\left.\overline{c}_{k_{z}''-q,\tau''}\left(\tau^{x}\sigma_{0}\right)c_{k_{z}''\tau''}\right]\right\rangle.
\end{multline}
Next, we use Wick theorem, leaving only connected diagrams. For a 4-operator expression it can be proven that
\begin{multline}
\left\langle T\left[c_{k_{1}}\left(\tau_{1}\right)\otimes\overline{c}_{k_{2}}\left(\tau_{2}\right)\overline{c}_{k_{z}'+q}\left(\tau'\right)\left(\tau^{x}\sigma_{0}\right)c_{k_{z}'}\left(\tau'\right)\right]\right\rangle \\
=\left\langle T\left[c_{k_{1}}\left(\tau_{1}\right)\otimes\overline{c}_{k_{z}'+q}\left(\tau'\right)\right]\right\rangle \left(\tau^{x}\sigma_{0}\right)\\
\times\left\langle T\left[c_{k_{z}'}\left(\tau'\right)\otimes\overline{c}_{k_{2}}\left(\tau_{2}\right)\right]\right\rangle .
\end{multline}
Similarly, to the second order we get
\begin{multline}
-\frac{\eta{}_{0}^{2}}{12}\sum_{k_{z}}\hat{G}_{k_{1}}\left(\tau_{1}-\tau'\right)\left(\tau^{x}\sigma_{0}\right)\hat{G}_{k_{z}}\left(\tau'-\tau''\right)\\
\times\left(\tau^{x}\sigma_{0}\right)\hat{G}_{k_{1}}\left(\tau_{2}-\tau''\right),
\end{multline}
where we used the relation $\left(\left|A\right\rangle \otimes\left\langle B\right|\right)\left\langle C|\tau^{x}\sigma_{0}|D\right \rangle =\left(\left|A\right\rangle \otimes\left\langle C\right|\right)\tau^{x}\sigma_{0}\left(\left|D\right\rangle \otimes\left\langle B\right|\right)$. Consequently, in the second order for the proper part we obtain
\begin{multline}
\hat{\Sigma}^{f}=-\frac{\eta{}_{0}^{2}}{12}\tau^{x}\sigma_{0}\sum_{\mathbf{k}_{\perp}k_{z}}\hat{G}_{\mathbf{k}_{\perp}k_{z}}\tau^{x}\sigma_{0}\\
=\Sigma_{0}^{f}+\Sigma_{\mu}^{f}\gamma_{\mu},
\end{multline}
where we restore the index $\mathbf{k}_{\perp}$ and the corresponding summation.

For simplicity, here we will restrict ourselves to the case without Zeeman splitting. Let us introduce the notation
\begin{gather}
\gamma_{1}=-\tau^{z}\sigma_{y},\gamma_{2}=\tau^{z}\sigma_{x},\\
\gamma_{3}=\tau^{x}\sigma_{0},\gamma_{4}=\tau^{y}\sigma_{0},
\end{gather}
then the Hamiltonian will be written in the form
\begin{multline}
\mathcal{\overline{H}}_{\mathbf{k}}=\gamma_{1}\upsilon_{F}p_{x}+\gamma_{2}\upsilon_{F}p_{y}+\gamma_{3}\left(\overline{\Delta}_{S}+\Delta_{D}\cos k_{z}d\right)\\
+\gamma_{4}\Delta_{D}\sin k_{z}d.
\end{multline}
It is easy to show that $\gamma_{i}^{2}=\mathbb{I}_{4\times4}$, $\gamma_{i}\gamma_{j}+\gamma_{j}\gamma_{i} =0$ for $i\neq j$. Then
\begin{gather}
\hat{G}_{\mathbf{k}_{\perp}k_{z}}=\frac{\varepsilon+\mathcal{\overline{H}}_{\mathbf{k}}}{\varepsilon^{2}-\upsilon_{F}^{2}k_{\perp}^{2}-\overline{\Delta}^{2}\left(k_{z}\right)}.
\end{gather}
Because the summation over impulses is carried out within symmetrical limits, then $\Sigma_{1}^{f}=\Sigma_{2}^{f}=\Sigma_{4}^{f}=0$. And for $\Sigma_{0}^{f}$ and $\Sigma_{3}^{f}$ we obtain
\begin{multline}
\Sigma_{0}^{f}\simeq-\frac{d_{\perp}^{2}\eta{}_{0}^{2}\varepsilon_{F}}{48\pi\upsilon_{F}^{2}\hbar^{2}}\\
\times\ln\left|\frac{\varepsilon_{F}^{2}-\left(\overline{\Delta}_{S}-\Delta_{D}\right)^{2}-\overline{\Delta}_{S}\Delta_{D}-\frac{\upsilon_{F}^{2}\hbar^{2}}{d_{\perp}^{2}}}{\varepsilon_{F}^{2}-\left(\overline{\Delta}_{S}-\Delta_{D}\right)^{2}-\overline{\Delta}_{S}\Delta_{D}}\right|,
\end{multline}
\begin{gather}
\Sigma_{3}^{f}=\frac{\overline{\Delta}_{S}+\frac{5}{6}\Delta_{D}}{\pi\varepsilon_{F}}\Sigma_{0}^{f}
\end{gather}
Note, that the exact calculation of these integrals is not difficult, but in this case cumbersome expressions arise, and therefore we expanded the integrands.

\bibliography{apssamp}
\end{document}